\documentclass[sigconf]{acmart}

\usepackage{enumitem}

\AtBeginDocument{%
  }

\copyrightyear{2026}
\acmYear{2026}
\setcopyright{cc}
\setcctype{by-nc-nd}
\acmConference[UIST '26]{The 39th Annual ACM Symposium on User Interface Software and Technology}{November 02--05, 2026}{Detroit, MI, USA}
\acmBooktitle{The 39th Annual ACM Symposium on User Interface Software and Technology (UIST '26), November 02--05, 2026, Detroit, MI, USA}
\acmDOI{10.1145/3830398.3830547}
\acmISBN{979-8-4007-2856-3/2026/11}

\newcommand{\paragraphB}[1]{\paragraph{{\textbf{#1. }}}}

\newcommand{\med}[1]{$\tilde{x} = #1$}

\newcommand{\toolname}{\textit{Mirror Skin}}

\begin{document}

\author{David Wagmann}
\affiliation{%
  \institution{Saarland University,\\ Saarland Informatics Campus}
  \city{Saarbrücken}
  \country{Germany}
}
\email{wagmann@cs.uni-saarland.de}

\author{Matti Krüger}
\affiliation{%
  \institution{Honda Research Institute Europe}
  \city{Offenbach/Main}
  \country{Germany}
}
\email{matti.krueger@honda-ri.de}

\author{Chao Wang}
\affiliation{%
  \institution{Honda Research Institute Europe}
  \city{Offenbach/Main}
  \country{Germany}
}
\email{chao.wang@honda-ri.de}

\author{Michael Gienger}
\affiliation{%
  \institution{Honda Research Institute Europe}
  \city{Offenbach/Main}
  \country{Germany}
}
\email{michael.gienger@honda-ri.de}

\author{Jürgen Steimle}
\affiliation{%
  \institution{Saarland University,\\ Saarland Informatics Campus}
  \city{Saarbrücken}
  \country{Germany}
}
\email{steimle@cs.uni-saarland.de}

\title[Mirror Skin]{Mirror Skin: In Situ Visualization of Robot Touch Intent on \\ Robot Skin}




%
% Keywords. The author(s) should pick words that accurately describe
% the work being presented. Separate the keywords with commas.


% A "teaser" image appears between the author and affiliation
% information and the body of the document, and typically spans the
% page.

\begin{teaserfigure}
  \includegraphics[width=\textwidth]{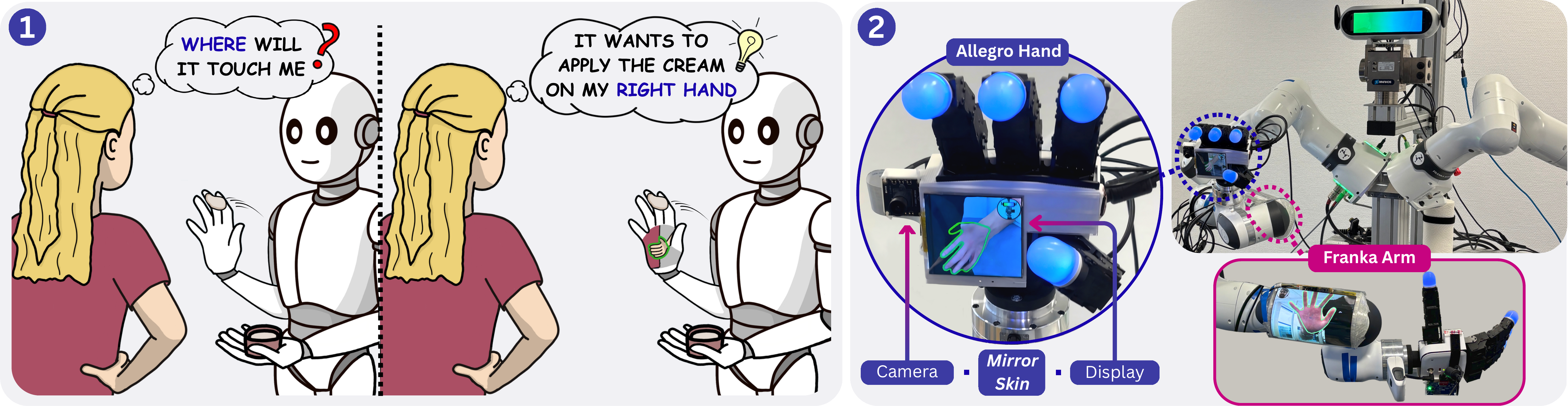}
  \caption{\toolname~communicates a robot’s touch intent through mirror‑like visual reflections on robot skin (1). We developed the \toolname~concept through a VR design exploration and built a proof‑of‑concept system (2). User studies in VR and with the physical prototype showed that \toolname~efficiently conveys robot touch intent and enhances the user experience.}
  \Description{Left: Shows two drawn illustrations that motivate touch intent communication. In the first one, a woman stands in front of a humanoid robot that has cream on its hand, which it apparently wants to apply to the woman. However, the woman seems confused about the robot's precise action. She seems reluctant and thinks, “Where is it going to touch me?” In the second illustration, the figure shows the same scenario, but now the robot displays mirror-like visual feedback of the woman's right hand on the robot’s approaching hand. The woman seems to understand and thinks, “It wants to apply the cream on my right hand.” Right: Images of our proof-of-concept system of Mirror Skin. First, one can see a robot with two arms and hands that seems to move its hand forward to touch someone, while Mirror Skin feedback is displayed on its hand. A first callout shows the Allegro hand of the robot and labels its two main components, that is, a camera and a display. In a second callout, one sees the Mirror Skin prototype on a Franka arm. More details about the hardware prototype are stated in Section 6.1.}
  \label{fig:teaser}
\end{teaserfigure}

\renewcommand{\shortauthors}{Wagmann et al.}
\begin{abstract}
Effective communication of robot touch intent is essential for safe and predictable physical human-robot interaction.
While intent communication has been widely studied, existing approaches lack the spatial specificity and semantic depth necessary to efficiently convey robot touch intent. 
We present \toolname, a cephalopod-inspired concept that mirrors in-situ visual representations of a human’s body parts onto the corresponding robot’s touch region to communicate \textit{who} shall initiate touch, \textit{where} it will occur, and \textit{when} it is imminent. 
We informed the design of \toolname~through a structured design exploration with experts and demonstrate the real-world feasibility of \toolname~with a proof-of-concept prototype. 
User studies in VR and with the physical prototype showed that \toolname~significantly improves accuracy and response times for interpreting touch intent and improves the user experience during physical human-robot interactions.
\end{abstract}

\maketitle

\section{Introduction}
A new generation of robots is emerging that engages in physical interaction with humans across various domains, including manufacturing~\cite{Matheson2019_HRI_Manufacturing, VW_2025}, healthcare~\cite{Speech_Touch_Feeling,Gonzalez_2021_social_robots_hospitals}, and social interactions~\cite{shiomi2020_survey_social_touch_HRI}. In such physical human-robot interactions (pHRI), it is important for humans to know when the robot intends to make contact \cite{pHRI_PreTouch}, as this awareness is fundamental for creating predictability~\cite{Speech_Touch_Feeling}, ensuring safety~\cite{Rubagotti2022_perceived_safety_phri}, and improving task performance \cite{AR_Trajectory1}. 

However, despite extensive research on the effects of robot-initiated touch in human-robot interaction (HRI), there remains a gap in understanding how to effectively and efficiently communicate a robot’s touch intent. While a substantial body of work addresses general robot intent communication, only few prior studies have included signaling of touch intent. These have predominantly relied on high-level verbal cues (notably,~\cite{Speech_Touch_Feeling, Speech_Pepper}), which may not always be favorable \cite{Speech_Touch_Feeling} or effective \cite{Speech_Gaze}, foregoing alternatives that may enhance clarity and user experience (UX).
This is problematic, as uncertainty about \textit{who} shall initiate touch and \textit{where} exactly it will occur can lead to expectation mismatches, potentially lowering the efficiency and safety of the interaction.
Therefore, we argue that it is important to explore dedicated communication concepts for human-robot touch to convey rich semantic feedback.

To address this gap, we draw inspiration from animals--such as cephalopods, lizards, and chameleons--that utilize visual color and texture changes as a powerful form of communication \cite{Animal_Cephalopod1, Animal_Chameleon2, Animal_Communication}. 
Cephalopods are particularly interesting due to their ability to alter skin color and texture with a high spatial and temporal resolution, for instance, to reflect parts of the environment onto their skin and to create visual highlights to direct attention toward, or away from, specific body regions \cite{Animal_Cephalopod2}.
These capabilities offer unique design opportunities for dynamic robot skin-based communication.

As a result, we present \toolname: a cephalopod-inspired concept that employs high-resolution visual feedback on the robot’s skin to communicate robot touch intent. Using the metaphor of a mirror, it communicates \textit{when} a touch event is imminent, \textit{who} shall initiate touch and \textit{where} it will occur. This is achieved by mirroring live reflections of the interacting human body part onto the surface of the robot’s touching body part (cf.,~\autoref{fig:teaser}). For instance, mirroring the human’s arm onto the robot’s moving end effector can indicate the robot’s intent to initiate contact at that body region. Conversely, reflecting the human’s hand onto the robot’s stationary arm can signal an invitation to touch the robot at that location.

We chose the mirror metaphor as a form of semantically meaningful visual feedback, since mirror reflections are a powerful mechanism for self-recognition \cite{Mirror_Self-Recognition}. Humans develop the innate ability to identify their own bodies and specific body parts in reflections from an early age \cite{Mirror_Efficiency}, making it an efficient and intuitive visual mechanism for body-centered communication \cite{MirrorEyes_Human}. 

To systematically develop our concept, we conducted a broad design exploration in virtual reality (VR).
We identified six key dimensions that shape the design of \toolname~and iteratively prototyped design variants to maximize visual clarity and salience. 
We then evaluated those generated design variants in an exploratory VR study (N = 7) with robotics and design experts and refined our design of \toolname.
Using our refined implementation, we conducted a lab study (N = 12) in VR to quantitatively and qualitatively assess its effectiveness for conveying touch intent before and during touch motions. 
The results demonstrate that \toolname~significantly improves the accuracy and response time for interpreting robot touch intent compared to a baseline condition consisting of robot gestures and gaze.
Finally, we implemented a physical proof-of-concept system incorporating high-resolution displays and wide-angle cameras to instantiate \toolname~in a real-world setting, enabling evaluation of its effects on intent communication as well as on pragmatic and hedonic UX. 
A lab study (N = 12) confirmed that compared to robot gestures, \toolname~communicated the robot's touch intent more clearly and yielded better pragmatic and hedonic UX, including higher confidence to engage in physical interaction with the robot.

In summary, this paper contributes:
\begin{enumerate}[noitemsep,topsep=0pt]
    \item \toolname, a cephalopod-inspired concept for semantically rich visual communication of touch-related intentions on robot skin, including \textit{who} shall initiate touch, \textit{when} touch is imminent and \textit{where} the touch is happening.
    \item An iterative design exploration that introduces and investigates six key components of \toolname, validated through an exploratory study with domain experts.
    \item Findings from a VR user study that validates the suitability of \toolname~for efficiently conveying robot touch intent.
    \item A physical proof‑of‑concept implementation of \textit{Mirror Skin} and real-world user study demonstrating improved UX.
\end{enumerate}
\section{Related Work}
This work is informed by prior work on pHRI and communication of robot intent in HCI and HRI.

\subsection{Physical Human-Robot Interaction}
Physical human-robot interaction has gained increasing attention due to the central role of touch in human social and cooperative interactions. Recent advances in robotic sensing technologies \cite{Sensing_Overview, Sensing_Skin}, electronic skin \cite{eSkin_AI, eSkin_Flexible, eSkin_Full} and autonomous behavior \cite{Robot_Autonomous_Street, Autonomy_Drones, AI_Trajectory} are enabling robots to engage in increasingly rich, physical interactions with humans: for example, to communicate emotions \cite{vanerp2013_how_to_touch_humans, willemse2017_responses_to_robot_initiated_touch}, or to collaborate with humans, such as in object handovers~\cite{gienger2018_cooperative_HRI_contact_changes, pHRI_PreTouch, Gaze_Handover}), and caregiving tasks~\cite{Gonzalez_2021_social_robots_hospitals, Speech_Touch_Feeling}. Conversely, humans may touch robots to express emotions~\cite{Emotions_2018}, provide guidance \cite{pHRI_Guidance1, pHRI_Guidance2}, or issue instructions.
A substantial body of work has investigated affective touch, focusing on both improving the tactile quality of robot touch (e.g., force or temperature \cite{pHRI_TouchQuality}) and understanding its psychological and behavioral effects on humans~\cite{vanerp2013_how_to_touch_humans, pHRI_CalmingEffects, pHRI_HealthcarePractitioners}. For instance, studies have shown that robot-initiated touch can foster comfort, increase trust, reduce stress, and enhance perceived social attributes, as well as influence the bonding with the robot~\cite{shiomi2020_survey_social_touch_HRI, willemse2017_responses_to_robot_initiated_touch, pHRI_CalmingEffects}.

\begin{sloppypar}
These findings demonstrate the broad potential of physical human-robot interaction in real-world scenarios. But, as robots become increasingly autonomous, driven by advances in AI, human-robot interactions become more dynamic. 
This raises the need for robots to communicate future touch intent to make touch actions interpretable, efficient and safe.
However, beyond verbal cues, which may not always be applicable~\cite{NonVerbal_Lemasurier} or effective \cite{Speech_Touch_Feeling, Speech_Gaze, Gestures_Efficiency}, there is currently a gap in research on how robots communicate touch intent. 
Parallel work has identified video-based visual feedback as a promising modality for conveying touch intent \cite{ReadyForTheTouch}. We advance this direction through~\toolname, which leverages the robot's skin as a mirror-like display to communicate \textit{who} shall initiate touch, \textit{where} it will occur, and \textit{when}
it is imminent.

\end{sloppypar}

\subsection{Non-Verbal Robot Intent Communication}

To convey robot intent, prior research has explored various forms of non-verbal feedback~\cite{Pascher_Intent}:

Haptic feedback has been applied to enhance the communication with tele-operated or mobile robots. However, it is typically limited to conveying predefined states or actions \cite{Haptic_Collision, Haptic_AR}, making the communication of richer semantic information challenging.

A very versatile method for conveying robot intent is feedback that the user can visually observe~\cite{NonVerbal_Lemasurier}.
Research extensively focused on motion-based cues such as eye, head, and arm movements \cite{Gesture_Finger, NonVerbal_Lemasurier, MirrorEyes_Object}.
In particular, robot gaze has been used to increase the clarity of robot actions \cite{NonVerbal_Lemasurier}, to improve spatial referencing of objects \cite{MirrorEyes_Object} and humans \cite{MirrorEyes_Human}, and to enhance performance in interactive tasks \cite{Gaze_Performance, Gaze_Handover}.
Moreover, motion trajectories have been optimized for legibility \cite{Motion1, Motion_Collaboration1}, enabling humans to rapidly infer the robot’s intended target.
Finally, robot gestures can enhance robot communication \cite{Gestures_Efficiency} and prior work explored how to design \cite{Gesture_Pointing, Gesture_Finger} or generate \cite{Gesture_Elicitation} them.

Other research investigated techniques for visual pre-motion intent communication that allow humans to anticipate the robot's actions early, even before the robot starts moving.
Here, light-based approaches have been explored to communicate robot state, motion direction, or intended actions on various types of robots \cite{NonVerbal_Lemasurier, Light_Bioluminescence, Light_Drones, Light_RLS}. Nevertheless, light cues have limited semantic capacity, and tasks such as conveying touch locations would require complex mappings, diminishing their effectiveness.
Another visual technique is projecting information onto the environment, which has been employed to communicate motion paths \cite{Projection_MotionPath, Projection_MotionPath2}, state \cite{Projection_State}, intentions \cite{Projection_Industry} and enhance spatial referencing of objects \cite{Projection_SpatialReference}. However, projection-based methods are constrained by lighting conditions and occlusion due to objects and other entities in the environment \cite{NonVerbal_Lemasurier}.
To mitigate these limitations, researchers have utilized augmented reality (AR). For instance to visualize motion direction \cite{AR_Arrow, AR_Drone} or trajectories \cite{AR_Trajectory1, AR_Drone}. Moreover, AR has been used to facilitate object handovers \cite{AR_Handover} and to display robot decision making \cite{Haptic_AR}. But a key limitation of AR is its reliance on additional hardware, restricting its applicability beyond dedicated settings.

Therefore, to convey richer semantic information without additional hardware requirements, one can provide feedback directly on the robot's body. Scholz et al. employed a flexible display wrapped around the robot’s arm, demonstrating the feasibility of high-fidelity, body-localized visual feedback \cite{VW_2025}. 
Extending this idea, we envision the robot’s skin as a high-resolution display for visual communication, moving beyond simple text or symbolic cues. Inspired by Krüger et al., who used virtual eyes as a mirror to reflect objects \cite{MirrorEyes_Object} or people \cite{MirrorEyes_Human} in order to improve spatial target identification, we leverage the robot’s skin as a dynamic mirror that selectively focuses on human body parts to communicate human-robot touch.

\subsection{Evaluating Human-Robot Interaction in VR}
VR and AR simulations are increasingly used in HRI research due to their effectiveness in evaluating novel robotic system designs through rapid prototyping~\cite{Prototyping_FakeIt, Pepper_Social}. Furthermore, prior work has shown that VR-based and physical robot prototypes yield comparable results for pre-touch aspects like visual feedback~\cite{VR_SocialArash, Prototyping_Rojas}.

Consequently, in this paper, we opt for a two-fold approach: First, to overcome hardware and scalability constraints, we conduct a broad exploration of the~\toolname~concept within a VR environment, alongside a deep investigation of its effectiveness in conveying robot pre-touch intent. Second, we evaluate the UX of~\toolname~during physical interactions using a proof‑of‑concept prototype in the real world.
\section{Mirror Skin}
In the following, we first explain the general concept, as well as a fundamental first version of \toolname. Afterwards, we present an iterative design exploration to improve upon the initial version.

\subsection{Concept}

\begin{figure} [t]
    \centering
    \includegraphics[width=0.97\linewidth, height=0.65\linewidth]{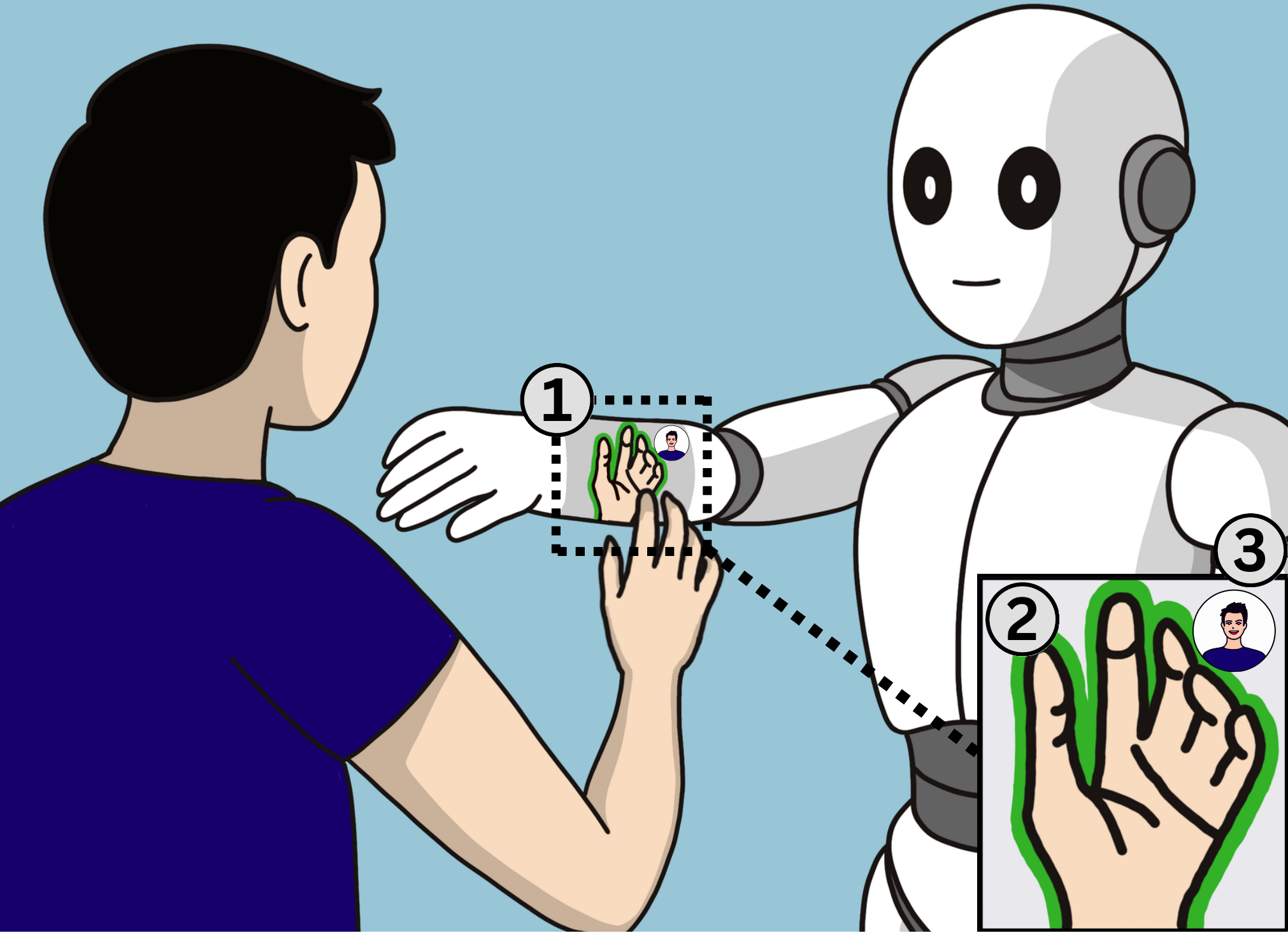}
    \vspace{-0.15cm}
    \caption{\toolname, conveys the contact locations of (1) the robot; (2) the human and (3) who shall initiate touch.}
    \Description{The figure shows an illustration of a robot using Mirror Skin to convey touch intent to a man standing in front of the robot. The robot has its right arm stretched out, offering it to the man and the man is moving its right hand towards the arm to touch it. On the right arm of the robot, one sees a mirror image of the man’s right hand and an icon of the man. Numbers in the illustration explain the components of the mirror skin concept. They show that 1) the location of the mirror image, that is, the robot’s right arm conveys the contact location of the robot; 2) the mirrored body part of the man, that is, his right hand conveys the contact location of the human and 3) the icon, that is, an image of the man denotes who shall initiate the touch.}
    \label{fig:concept}
    \vspace{-0.4cm}
\end{figure}

Similar to cephalopods, we utilize the robot's skin as a high-resolution display conveying visual information.
In order to leverage this display for touch intent communication, we must visually convey the three key aspects of the touch interaction:
(1) the human body part involved in the touch action, (2) the robot body part involved in the touch action and (3) who shall initiate the touch action.

To encode the location of the intended touch actions, we transform the robot’s skin at the location that is involved in the touch interaction into a visual "mirror". This mirror dynamically displays the targeted human body part and its surrounding environment. The mirror reflection continuously follows the motion of the targeted body region, thereby establishing a spatially and temporally aligned one-to-one mapping between the human and robot body.
This mirror metaphor leverages humans’ familiarity with reflective feedback, which enhances self-identification, and provides an intuitive reference for body-centered interactions \cite{Mirror_Efficiency, Mirror_Self-Recognition, MirrorEyes_Human}.
As a result, it enables humans to infer both the occurrence and precise location of the anticipated physical interaction (cf., \autoref{fig:concept}).
We selected the body-localized approach over a centralized display (e.g., Pepper's tablet \cite{Screens_Body}) because this directly encodes the robot's touching body part without requiring an additional mapping. At the same time, it reduces the need to divide attention between the robot’s touching body part and the provided visual feedback.

To communicate who shall initiate touch pre-motion, we augment the reflection with actor-specific portraits: A human portrait denotes that the human is expected to touch the robot, whereas a robot portrait indicates that the robot intends to initiate contact.

\begin{figure*}[t]
    \centering
    \includegraphics[width=0.92\linewidth, height=0.64\linewidth]{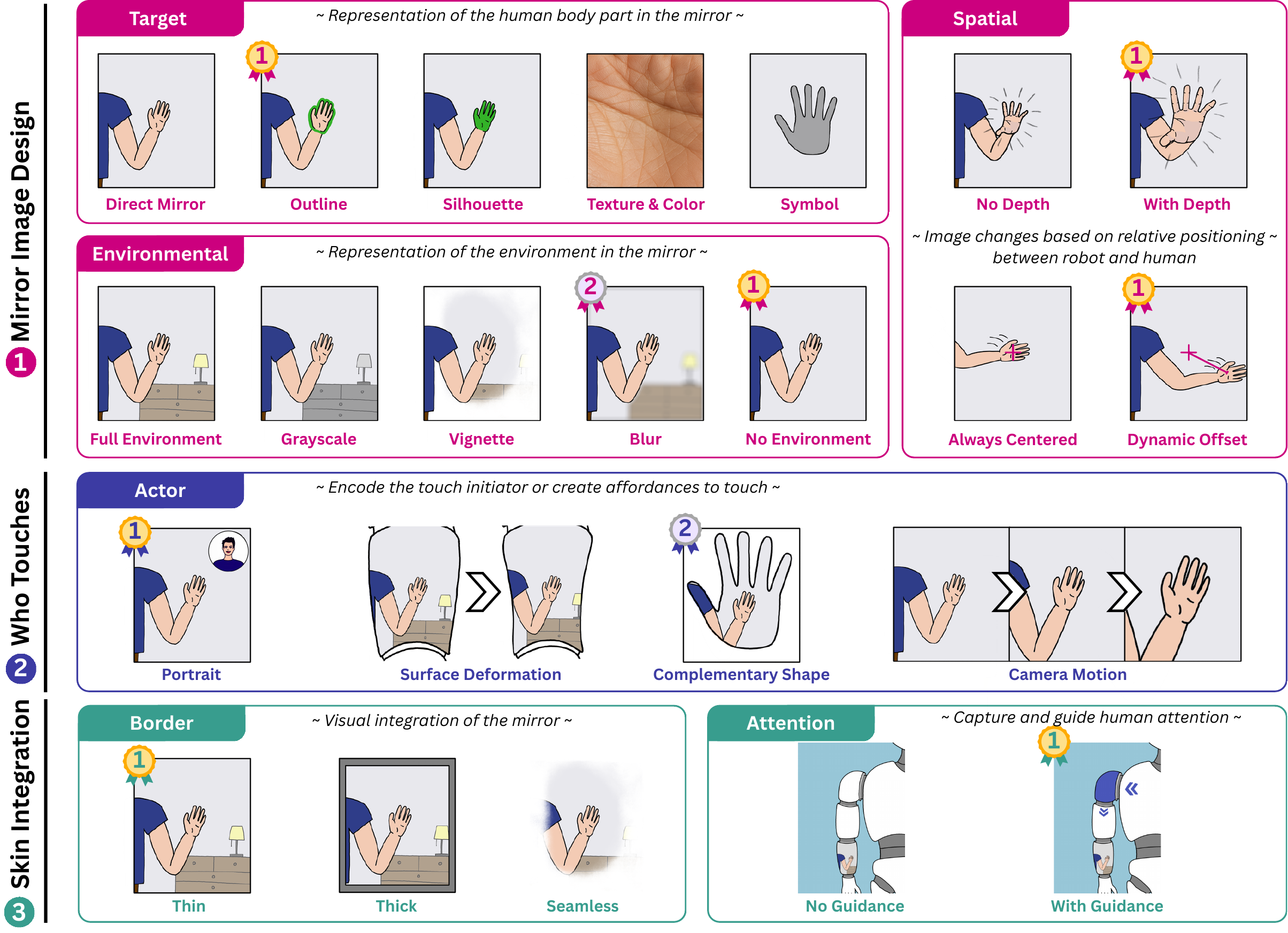}
    \caption{We investigated six dimensions for visually encoding touch intent on robot skin. For each dimension, we explored multiple design variants to enhance \toolname\textit{'s} clarity and salience and utilize the top-ranked variants for the final design.}
    \Description{This figure provides a comprehensive overview of the six encoding dimensions that constitute the design of Mirror Skin. In each dimension, we show a sketch that visualizes every design variant that we explored and described in our structured design exploration (see Section 3.2). The sketch shows the reflection of a human right shoulder to hand with some furniture in the background. The different encodings change this image accordingly, for example, blurring the furniture in the background or providing a green outline around the targeted hand.}
    \vspace{-0.1cm}
    \label{fig:designSpace}
\end{figure*}

\subsection{Structured Design Exploration}
Given that the design space for skin-based communication of robots remains largely unexplored, we performed a structured design exploration in VR to improve the fundamental version of \toolname~regarding clarity~\cite{Ambiguity_Reduction1} and salience~\cite{Saliency_Increase1}.
To identify the key parameters that shape the design of \toolname~we first subdivided the visual feedback of \toolname~into six dimensions (cf., \autoref{fig:designSpace}). 
We then iteratively prototyped within each dimension, generating improved variants informed by prior work and author discussions.

First, to improve the salience and clarity of the content shown in the mirror reflection, we structured the information of the mirror into three dimensions (cf., \autoref{fig:designSpace}$~\rightarrow~$Mirror Image Design):\\

\noindent\textbf{\textit{Target encoding}} specifies how the human body part involved in the touch interaction is represented on the robot’s skin. To provide an unambiguous encoding while minimizing cognitive load, we created several design variants that differ in their degree of abstraction:

\begin{itemize}[leftmargin=*, noitemsep, topsep=0pt]
    \item[] \textsc{\textbf{Direct Mirror:}} Our initial version of \toolname~employs a direct mirror image of the human body part.
    \item[] \textsc{\textbf{Outline \& Silhouette:}} To enhance salience and granularity within the live image, we follow insights from target identification research \cite{VR_TargetIdentification, Silhouette_2008} and apply visual highlighting in the form of an outline or silhouette to the target in the direct mirror.
    \item[] \textsc{\textbf{Texture \& Color:}} Inspired by cephalopod camouflage \cite{Animal_Cephalopod1, Animal_Cephalopod2} and color communication in HRI \cite{Light_Bioluminescence}, we sample the surface texture and color of the target and map it onto the robot’s skin.
    \item[] \textsc{\textbf{Symbol:}} Symbols are an efficient communication modality employed in existing robotic systems \cite{VW_2025}.
\end{itemize}

\noindent\textbf{\textit{Environmental encoding}} involves the visual representation of the background in the mirrored image. Although environmental cues can facilitate spatial disambiguation by adding context, visual clutter can negatively impact target identification.
Thus, we investigated the following encodings, which gradually reduce the provided environmental information:

\begin{itemize}[leftmargin=*, noitemsep, topsep=0pt]
    \item[] \textsc{\textbf{Full Environment:}} Our initial version of \toolname~mirrors the full environment onto the robot's skin.
    \item[] \textsc{\textbf{Grayscale:}} The environment is desaturated to lower its visual salience while maintaining spatial information.  
    \item[] \textsc{\textbf{Vignette:}} The environment fades out gradually toward the periphery to guide attention to the central target region.  
    \item[] \textsc{\textbf{Blur:}} The environment is blurred to reduce visual detail while preserving overall scene structure.
    \item[] \textsc{\textbf{No Environment:}} The environment is entirely removed to focus solely on the human body.  
\end{itemize}

\noindent\textbf{\textit{Spatial encoding}} specifies how the mirror-image changes based on relative positioning and movement between the human and the robot. By adding spatial cues, we simulate mirror-like behavior, thereby increasing the human's spatial awareness.
We explored the following spatial encoding strategies:

\begin{itemize}[leftmargin=*, noitemsep, topsep=0pt]
    \item[] \textsc{\textbf{Depth:}} The reflection can either maintain a constant size regardless of distance (no depth) or vary in size relative to the distance between the human body part and the robot’s surface (with depth), analogous to a real mirror.  
    \item[] \textsc{\textbf{Dynamic Offset:}} Instead of remaining fixed at the mirror's center, the reflection is temporarily displaced along the mirror plane in response to human movement and gradually re-centers. This creates the impression that the reflection follows the human, approximating the behavior of a physical mirror. \\
\end{itemize}

\noindent
Second, to convey who shall initiate touch through the mirror, we investigate alternative representations for encoding the initiating actor prior to any robot motion (cf., \autoref{fig:designSpace}$~\rightarrow~$Who Touches):\\

\noindent\textbf{\textit{Actor encoding}} differentiates between the \textit{intention} to touch the human and the \textit{invitation} to be touched by the human. Conveying this distinction clearly can reduce ambiguity in role allocation. We investigated the following encoding strategies:

\begin{itemize}[leftmargin=*,noitemsep, topsep=0pt] 
    \item[] \textsc{\textbf{Portrait:}} We explored portraits of varying fidelity representing either the human or the robot to indicate the initiator.  
    \item[] \textsc{\textbf{Surface Deformation:}} Inspired by prior research~\cite{ShapeChange_Exploration, Shape_GooY}, we visualize deformations on the robot's surface similar to how skin gets deformed when touched (i.e., indentations at the front to afford poking, or at the side to afford grasping).  
    \item[] \textsc{\textbf{Complementary Shape:}} By altering the shape of the projected area to match the human body part, e.g., a hand-shaped mirror, we create the visual affordance to align the human's hand with the displayed geometry, conveying an invitation.
    \item[] \textsc{\textbf{Camera Motion:}} Controlled zoom-in or zoom-out effects to signal whether the robot intends to approach the human or expects the human to initiate contact. \\
\end{itemize}

\noindent
Finally, we sought to explore how to visually integrate \toolname~on the robot's skin to make the feedback more noticeable. Thus, we explored  (cf., \autoref{fig:designSpace}$~\rightarrow~$Skin Integration):\\

\noindent\textbf{\textit{Border encoding}} determines the visual transition between the reflection and the robot's body. We explored the following variants:

\begin{itemize}[leftmargin=*,noitemsep, topsep=0pt]
    \item[] \textsc{\textbf{Thin:}} The initial version of \toolname~simply overlays the mirror image onto the skin of the robot, creating a thin transition between the mirror and the remaining robot skin.
    \item[] \textsc{\textbf{Thick:}} Adding a thick border around the reflection creates a distinct visual boundary, increasing noticeability but reducing the sense of integration with the robot’s body.
    \item[] \textsc{\textbf{Seamless:}} Seamless embedding of \toolname~into the robot’s body surface can strengthen the association between the visual feedback and the robot itself. However, increased visual integration may reduce the salience of the reflection.
\end{itemize}

\noindent\textbf{\textit{Attention encoding}} aims to catch and guide the attention of the human towards the feedback of \toolname. This can be particularly helpful when feedback is presented on small robot parts (e.g., fingers), where visual cues may be easily overlooked.
To assess the role of attentional guidance, we compare:

\begin{itemize}[leftmargin=*,noitemsep, topsep=0pt] 
    \item[] \textsc{\textbf{No Guidance:}} The initial version of \toolname~relied solely on the salience of the mirror itself to catch the user's attention.  
    \item[] \textsc{\textbf{Visual Guidance:}} The robot provides a salient color cue at a visually prominent region of its body (e.g., the chest) to attract attention. The cue then propagates towards the robot part with \toolname~feedback, guiding the human’s focus to the mirror.  
\end{itemize}
\section{Exploratory Design Study} \label{sec:exploratory_study}
To evaluate the design variants regarding clarity and salience and to identify further opportunities for improvement, we conducted an exploratory study with robotics and design experts. These insights informed the refined design of \toolname.

\subsection{Method}

\paragraphB{Apparatus}
To rapidly iterate on the visual encoding strategies, we developed an interactive VR design environment coupled with a GUI.
The virtual environment was created with Unity 3D for the Meta Quest 3. We included both a male and female avatar to increase the participants' self-identification with their avatar, which is critical for attributing the mirrored feedback to one's body~\cite{Martin_Redirection}. To meet the quality standards for VR research, we followed previous experiments and used fully rigged avatars from the Microsoft RocketBox library~\cite{gonzalez-franco_rocketbox_2020} and FinalIK for realistic character animations. 

\paragraphB{Experimental protocol}
Following an initial calibration procedure to align the avatar with the user’s proportions, participants were positioned in front of a humanoid robot (cf.,~\cite{Pepper_Social, Pepper_Teleoperation}) with integrated \toolname. 
Our setup supported real-time manipulation of all six encoding dimensions, allowing users to select and combine them interactively and observe their effects on the robot’s skin.
Furthermore, the environment enabled specification of discrete touch locations and provided predefined poses and motion sequences for both intention and invitation cues, allowing them to examine \toolname~across varying surface geometries and interaction scenarios.

We first introduced the participants to the different encoding dimensions and our proposed designs. Afterwards, they explored and configured their preferred combination of encoding instantiations and discussed their design choice. We recorded all selections in VR; during the whole experiment, the participants were asked to think aloud while the experimenter took notes.
We concluded with a semi-structured interview to discuss participants’ final configurations and to investigate further opportunities for improvement.
The study took approximately 60 minutes and was audio-recorded.

\paragraphB{Participants}
We recruited 7 participants (aged 28 to 34, $\bar{x} = 30.71$; 5 male, 2 female) with normal or corrected-to-normal vision. 
Two of them were roboticists with expertise (>5 years) in HRI and human-robot collaboration; two were professional designers (>9 years) with backgrounds in user interface, UX, and interaction design; and the remaining three participants had interdisciplinary expertise (>5 years), combining knowledge from HRI and HCI.

\subsection{Results}
In the following, we present the expert opinions on our designs and how we refined \toolname~accordingly.

\paragraphB{Target encoding}
Participants emphasized the importance of clearly conveying the targeted human body part. In that context, the \textsc{\textbf{Direct Mirror}} was not considered clear enough, because \textit{``it’s very ambiguous [...] what [the robot] is actually looking at''} (P6).
Consequently, all participants agreed that the target encoding should be made more salient and identified highlighting (i.e., \textsc{\textbf{Outline}} or \textsc{\textbf{Silhouette}}) as the most effective approach for the target representation. Moreover, they all favored the \textsc{\textbf{Outline}}, because it retains more detail of the target than the \textsc{\textbf{Silhouette}}.
While \textsc{\textbf{Symbols}} were generally well-received, they lack the dynamic interaction that is available through the mirror: \textit{``It is very static and there is no kind of interaction''} (P4). Additionally, P1 mentioned that \textit{``it's also a bit more personal to have my moving body displayed here''}, which facilitates the identification between the projected and real body part.
Lastly, all participants disliked the plain \textsc{\textbf{Texture \& Color}}, as they are too abstract and ambiguous.

As a result, we chose a live mirror image with an additional \textsc{\textbf{Outline}} as our target encoding for \toolname.

\paragraphB{Environmental encoding}
All experts agreed that it is important to reduce the provided background details to avoid visual clutter and they distinguished between two use cases: 
If it is \textit{``important to interact with objects and environment [...] having these slight environmental cues still helps''} (P1) and therefore, participants chose the \textsc{\textbf{Blur}} feature, as \textit{``blur is suited for blending information without removing it''} (P2)
Otherwise, if the robot is solely focused on the human, then it is optimal to remove the background completely (i.e., \textsc{\textbf{No Environment}}), because \textit{``then there's much clearer focus [...] that it's about my body''} (P7).

Since the present implementation of \toolname~is designed to facilitate human-robot touch in single-user scenarios without involving environmental objects, we utilize \textsc{\textbf{No Environment}}, but emphasize the usefulness of \textsc{\textbf{Blur}} for multi-target scenarios.

\paragraphB{Spatial encoding}
All participants agreed that including \textsc{\textbf{Depth}} is a good idea, as it resembles the physical property of mirrors and thus helps establish the link between the reflection and our body.
The \textsc{\textbf{Dynamic Offset}} feature was mostly preferred (P2, P3, P5-P7), because it contributes to the physical mirror feeling and similarly, P2 stated: \textit{``That looks more organic [...] feels less like tracking''}.
Nevertheless, participants (P1, P5, P7) suggested removing the spatial information for body parts with a small form factor like the fingertip, to increase the visibility during motion.

Accordingly, the final version of \toolname~incorporates both \textsc{\textbf{Depth}} and \textsc{\textbf{Dynamic Offset}} motion cues for interactions involving larger surface areas, whereas these cues are omitted for small body parts, such as the finger.

\paragraphB{Actor encoding}
Although approaches like \textsc{\textbf{Surface Deformation}} and \textsc{\textbf{Camera Motion}} were described as \textit{``interesting''} (P6), they were not regarded as intuitive, salient, or efficient to communicate who is going to act.
Generally, experts liked the \textsc{\textbf{Complementary Shape}} (P1, P3, P6-P7), which was considered the most intuitive, naturally inviting humans to initiate touch. But \textsc{\textbf{Portraits}} (P1-P7) were clearly regarded as the most efficient.
To enhance the interpretability of the portrait, experts (P1, P2, P6) emphasized that it is best to use real headshots of the robot and human to establish a direct connection between the portrait and the actors, similar to a second mirror. That way, the feedback can also convey the actor more easily in multi-user scenarios.

Due to these benefits, we implemented the headshot version of the \textsc{\textbf{Portrait}} for \toolname~and suggest \textsc{\textbf{Complementary Shapes}} as a tool to enhance interpretability during first-time interactions with the robot.

\paragraphB{Border encoding}
Experts expressed that the border encoding is less critical to its functional design compared to other encoding strategies, as it primarily influences visual aesthetics without substantially enhancing perceptual clarity or salience. 

Therefore, we retained the original \textsc{\textbf{Thin}} border, which maximizes available screen space relative to the other design alternatives.

\paragraphB{Attention encoding}
All experts supported the inclusion of an attention-grabbing mechanism (i.e., \textsc{\textbf{Visual Guidance}}). P4 mentioned: \textit{``I think it's cool, especially for the small fingertip because otherwise you wouldn't realize that something is happening there''}.
Participants also provided suggestions for improving the implemented version. These included changing the color to be less discouraging for making contact (e.g., blue \cite{Light_Bioluminescence}) (P4) and increasing its pace (P3, P5) to speed up the interaction.

Consequently, we incorporated a fast, uniform blue attention cue for \textsc{\textbf{Visual Guidance}}.

\section{Performance Study} \label{sec:efficiency_study}
To empirically validate the performance of \toolname~for communicating robot touch intent across various parts of the upper limbs, we built a VR environment and performed a controlled user study.

\begin{figure*}[t]
    \centering
    \includegraphics[width=1\linewidth, height=0.3\linewidth]{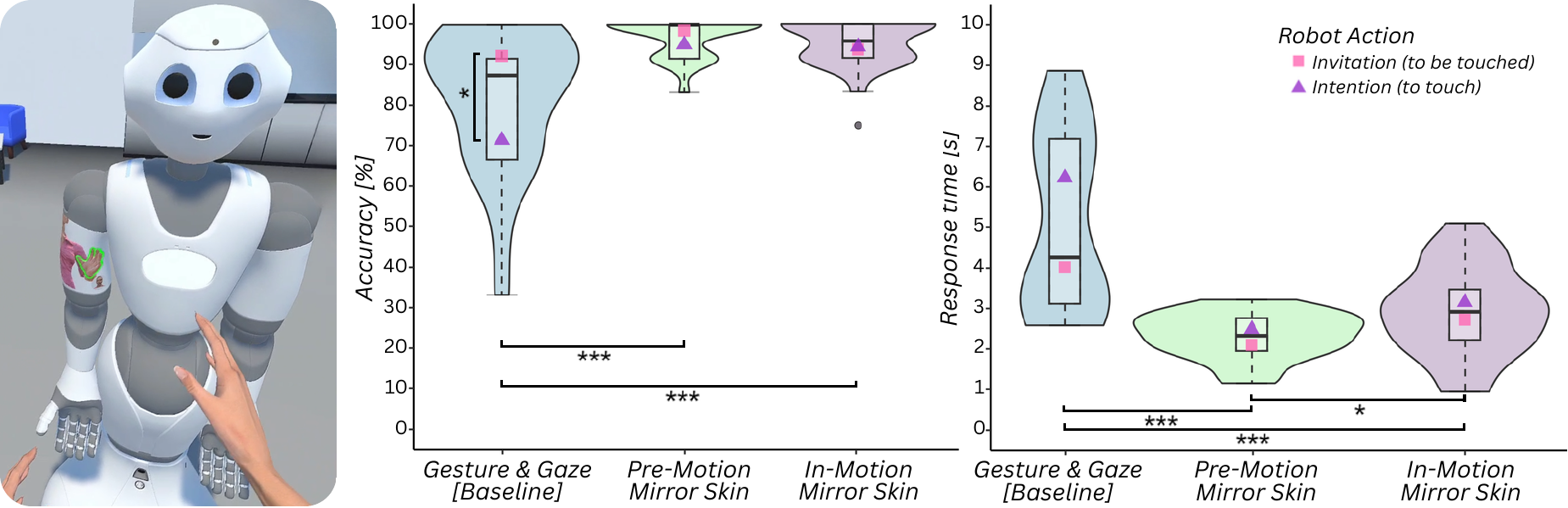}
    \caption{We investigated \toolname\textit{'s} performance for communicating robot touch intent (Left). Our results show that \toolname~cues before (\textit{pre-motion}) and during (\textit{in-motion}) robot touch motions both significantly enhance accuracy (Middle) and lower response times (Right) of touch intent interpretation, compared to the gesture \& gaze baseline.}    
    \Description{Left: An image from the VR environment. The image is from the first-person view of a human, embodying a female avatar that stands in front of a humanoid robot. The female avatar has her hands raised in front of her and one can see a mirror skin reflection of her right hand on the right upper arm of the robot. This demonstrates what Mirror Skin looked like during our VR studies. Middle: Superimposed violin- and boxplots: The x-axis distinguishes between the gesture & gaze (baseline), pre-motion Mirror Skin and in-motion Mirror Skin. The y-axis shows the measured accuracy of participants in percent  (0 to 100). The violin plots show that both Mirror Skin conditions significantly improved the accuracy compared to the gestures and gaze baseline. Precise value and analysis are stated in the main text. Right: Shows the same coordinate system as in the middle, but the y-axis indicates the response time in seconds (0 to 10). The violin plots show that both Mirror Skin conditions significantly improved the response time compared to the gestures and gaze baseline. Details and analysis are stated in the main text.}
    \label{fig:accuracy_responsetime}
\end{figure*}

\paragraphB{Experimental design and task}
For our study, we used the VR environment described in Section \ref{sec:exploratory_study}. Participants faced a robot that communicated imminent touch actions, and their task was to infer as quickly as possible \textit{who} should initiate touch and \textit{where} the touch should occur. 
After pressing a button on the VR controller, which stopped the timer and suppressed all robot feedback to prevent premature responses, participants answered verbally.
We restricted interactions to hand-initiated touch, given its predominance in human-robot touch and randomly chose the left or right hand with equal frequency. Likewise, the targeted body part--palm, forearm, or upper arm on either side--was randomly assigned on each trial.
We compared three visual communication strategies (cf.,~\autoref{sec:appendix_feedback}):\\
First, we investigated \textbf{\toolname}~for \textbf{pre-motion touch intent} communication, meaning that the robot remained in its default pose and communicated intent only through \toolname.\\
Second, we investigated \textbf{\toolname}~for \textbf{in-motion touch intent} communication to assess its performance during dynamic interactions. The robot either offered a body part for contact or approached the participant with the hand to signal touch intent. All approach trajectories followed a direct path to the participant’s body part, with motion and hand orientation dynamically adapting to the participant’s position. The robot always stopped prior to physical contact, ensuring that interactions remained strictly pre-touch. \\
Finally, to compare \toolname~to an established visual communication method (cf., \cite{Pascher_Intent, NonVerbal_Lemasurier}), we implemented a baseline condition consisting of the robot motion \textbf{gestures} from the previous condition, augmented with \textbf{robot gaze} that continuously tracked the relevant human body part to convey explicitly which body part should be involved in the interaction.

\paragraphB{Experimental variables}
Our study follows a within-subjects design in which each participant interpreted the robots' communicated touch actions across different types of feedback and action cues. Therefore, we consider two independent variables (IVs):
\begin{itemize}[leftmargin=*]
    \item \textsc{Feedback}: The robot communicates imminent touch actions using either gesture \& gaze [\textsc{Baseline}], pre-motion \toolname~[\textsc{Pre-M. MS}], or in-motion \toolname~[\textsc{In-M. MS}] feedback.
    \item \textsc{Action}: The robot can either express the \textsc{Intention} to touch the human or offer an \textsc{Invitation} to be touched.
\end{itemize}

Each participant completed 12 trials for each \textsc{Action} in randomized order, yielding 24 samples per \textsc{Feedback} condition
We counterbalanced the conditions for \textsc{Feedback} with a balanced Latin square, resulting in 3 x 24 = 72 samples per participant.
For each sample, we measured the following dependent variables (DVs):
\begin{itemize}[leftmargin=*]
    \item \textsc{Accuracy:} The percentage of correctly answered trials.
    \item \textsc{Response Time:} How fast the human identified the robot's intent.
\end{itemize}

\paragraphB{Procedure}
After filling out a consent and demographics form, participants went through the same calibration step as in the exploratory study. To ensure that participants could accurately associate their mirrored body parts with their avatar representation, we added a familiarization phase (cf.,~\cite{Martin_Redirection}), in which participants performed announced and free movements in front of a mirror.

During the study, each \textsc{Feedback} condition was introduced through a conceptual explanation followed by open-ended practice rounds.
During the trials, our system tracked participants' response times and the experimenter documented their answers.
After each level of \textsc{Feedback}, participants completed the NASA-TLX~\cite{NASA_TLX} to capture their task load, and rated how sufficient and efficient the feedback conveyed the touch intent on 7-point Likert scales.

We concluded with a semi-structured interview for more in-depth insights regarding participants' opinions and expectations.
The study session took approximately 60 minutes and the final interview was audio-recorded.

\paragraphB{Participants}
We recruited 12 participants (aged 22 to 31, $\bar{x} = 25.16$; 7 male, 5 female) with normal or corrected-to-normal vision and little (4/12) or no experience with HRI (8/12).

\paragraphB{Data analysis}
We tested the normality assumption of our data with Shapiro-Wilk tests and QQ-plots. Since normality was violated, we applied non-parametric tests to analyze our data:
First, we applied the Aligned Rank Transformation~(ART) repeated measures ANOVA as proposed by Wobbrock et al.~\cite{wobbrock_2011} to investigate interaction and main effects between \textsc{Feedback} and \textsc{Action} on accuracy and response time. For significant results, we conducted post hoc analyses using the ART-C procedure, as suggested by Elkin et al.~\cite{elkin_2021}. 
Effect sizes for ART were reported using partial eta-squared ($\eta_p^2$), classified as small~(>~.01), medium (>~.06), or large (>~.14)~\cite{Cohen1988}. For ART-C, we reported Cohen’s $d$, and classified it as small (>~.20), medium (>~.50), or large (>~.80)~\cite{Cohen1988}. Additionally, to analyze the effect of \textsc{Feedback} on task load and UX, we applied Friedman tests, followed by pairwise Wilcoxon signed-rank tests for significant results. We reported Kendall's $W$ and the rank-biserial $r$ as the measures of effect size and used Cohen's suggestions~\cite{Cohen1988} to classify them as small~(>~.10), medium~(>~.30), or large~(>~.50).

\subsection{Results}
\paragraphB{Accuracy}
Overall, the participants had very high accuracy rates (cf., \autoref{fig:accuracy_responsetime} (middle)), but scores varied between the \textbf{\textsc{Baseline} (81.93\%)}, \textbf{\textsc{Pre-M. MS} (96.86\%)} and \textbf{\textsc{In-M. MS} (94.10\%)}, showing that \toolname~supported interpreting the robot's intent. For deeper investigation, we performed an ART that revealed a significant main effect of \textsc{Feedback} on accuracy ($F_{2, 55} = 18.38,~p < .001,~\eta_p^2 = 0.40$). Post-hoc ART-C pairwise comparisons showed a significantly higher accuracy score for \textsc{Pre-M. MS} compared to \textsc{Baseline}~($p < .001,~d = 1.654$) and \textsc{In-M. MS} also yielded significantly higher accuracy scores than \textsc{Baseline}~($p < .001,~d = 1.323$). However, we found no significant difference between the accuracy scores of \textsc{Pre-M. MS} and \textsc{In-M. MS} ($p > .05$).
Next, the ART revealed a significant main effect of \textsc{Action} on accuracy, with \textsc{Invitation} having a significantly higher accuracy compared to \textsc{Intention} ($F_{1, 55} = 37.52,~p < .001,~\eta_p^2 = 0.41$).
Additionally, we also found a significant interaction effect between \textsc{Feedback} and \textsc{Action}~($F_{2, 55} =  11.27,~p < .001,~\eta_p^2 = 0.29$).
Post-hoc comparisons revealed a significantly higher accuracy for \textsc{Invitation} compared to \textsc{Intention} for the \textsc{Baseline} ($p < .05,~d = 1.392$), but showed no significant difference for \textsc{Pre-M. MS} and \textsc{In-M. MS} ($p > .05$). 
This observation is supported by post-hoc tests for \textsc{Intention}, which revealed significantly higher accuracy for \textsc{Pre-M. MS} compared to \textsc{Baseline}~($p < .001,~d = 1.881$) and for \textsc{In-M. MS} compared to \textsc{Baseline}~($p < .001,~d = 1.933$).
Finally, post-hoc tests for \textsc{Invitation} indicated a significantly higher accuracy for \textsc{Pre-M. MS} compared to \textsc{Baseline}~($p < .05,~d = 1.217$), but not between the remaining groups ($p > .05$).

\paragraphB{Response time}
All types of \textsc{Feedback} conveyed the intent within a reasonable time (cf., \autoref{fig:accuracy_responsetime} (right)); however, with different average response times between \textbf{\textsc{Baseline} (5.12s)}, \textbf{\textsc{Pre-M. MS} (2.29s)} and \textbf{\textsc{In-M. MS} (2.93s)}.
The ART revealed a significant main effect of \textsc{Feedback} on response time ($F_{2, 52} = 34.93,~p < .001,~\eta_p^2 = 0.57$). Post-hoc comparisons showed significantly faster response times of \textsc{Pre-M. MS} compared to \textsc{Baseline}~($p < .001,~d = -2.433$) and \textsc{In-M. MS} also yielded significantly faster response times than \textsc{Baseline}~($p < .001,~d = -1.642$). Additionally, we found significantly faster response times of \textsc{Pre-M. MS} compared to \textsc{In-M. MS} ($p < .05,~d = -0.791$). 
Next, we found a significant main effect of \textsc{Action} on response time, with \textsc{Invitation} having a significantly faster response time compared to \textsc{Intention} ($F_{1, 52} = 28.17,~p < .001,~\eta_p^2 = 0.35$).
Moreover, we also found a significant interaction effect between \textsc{Feedback} and \textsc{Action}~($F_{2, 52} =  8.25,~p < .001,~\eta_p^2 = 0.24$).
Contrary to the accuracy, post-hoc tests found no significant differences in response time for \textsc{Invitation} compared to \textsc{Intention} in any \textit{Feedback} condition ($p > .05$).
Furthermore, post-hoc comparisons for \textsc{Intention} revealed significantly faster response times for \textsc{Pre-M. MS} compared to \textsc{Baseline}~($p < .001,~d = -2.911$) and \textsc{In-M. MS} also showed significantly faster response times than \textsc{Baseline}~($p < .01,~d = -1.747$).
Lastly, post-hoc tests for \textsc{Invitation} indicated significantly faster response times for \textsc{Pre-M. MS} compared to \textsc{Baseline}~($p < .001,~d = -2.662$) and for \textsc{In-M. MS} compared to \textsc{Baseline}~($p < .01,~d = -1.641$).
These results demonstrate that \toolname~facilitates faster human comprehension of the robot's intended actions in both pre-motion and in-motion scenarios. 

\paragraphB{Task load}
The task load was similar across \textsc{Baseline} ($\bar{x} = 37.08$), \textsc{Pre-M. MS} ($\bar{x} = 35.76$) and \textsc{In-M. MS} ($\bar{x} = 38.47$) and a Friedman test did not find a significant effect of \textsc{Feedback} ($p > .05$).

\paragraphB{Usability \& User Opinions}
First, participants reported that the feedback was sufficient in every condition (\textsc{Baseline}: \med{6}; \textsc{Pre-M. MS}: \med{6.5}; \textsc{In-M. MS}: \med{7}) and a Friedman test showed no significant differences ($p > .05$), indicating that the task was comparably feasible with all \textsc{Feedback} types.
In line with the prior analysis of response time, participants rated \textsc{Pre-M. MS} (\med{6}) and \textsc{In-M. MS} (\med{6}) as more efficient for intent communication than \textsc{Baseline} (\med{4.5}). A Friedman test revealed a significant effect of \textsc{Feedback} on the perceived efficiency ($\chi^2(2) = 8.33,~p < .05,~W = 0.347$).
Post-hoc tests showed that participants regarded \textsc{Pre-M. MS} as significantly more efficient than \textsc{Baseline} ($p < .05,~r = 0.79$), but they found no significant differences between the remaining groups ($p > .05$).

In the follow-up interviews, participants indicated multiple reasons for the improved efficiency. First, several participants (P5, P6, P8, P9, P12) attributed this to the fact that \textit{``[one] didn’t have to wait for [the robot's] movement and gaze''} (P9), as all necessary information was immediately available.

At the same time, we observed participants independently leveraging the spatial-referencing advantages of \toolname~by moving their wrists or fingers to make the mirror feedback more effective (P1, P3-P10, P12). This is notable, as it suggests that the mirroring property offers advantages over static imagery in visual feedback.

Furthermore, all participants (P1-P12) mentioned that when the robot issued an \textsc{Intention}, using \textsc{Baseline} feedback, it was difficult to distinguish between adjacent body parts while the robot's hand was still further away. In contrast, this issue did not occur in the \textsc{Pre-M. MS} or \textsc{In-M. MS} conditions, making \toolname~particularly relevant for robot-initiated touch.

But despite the benefits of \toolname, multiple participants (P3-P7, P12) stated that it was more challenging to follow the feedback of \toolname~during arm motions.
This difficulty was repeatedly attributed to self-occlusion (P3, P5, P12), i.e., the robot blocks the line of sight towards the mirror with its own body, and to distortion (P6, P7, P12) caused by motion across curved surfaces, which impaired the clarity and interpretability of the mirrored feedback.
\section{Physical Prototype and User Experience Study}
To demonstrate the feasibility of \toolname~on a real robot and to investigate its effects on the UX, we implemented a proof-of-concept \toolname~prototype and conducted a user study.

\subsection{Proof-Of-Concept Prototype}

\begin{figure*}[t]
    \centering
    \includegraphics[width=1\linewidth]{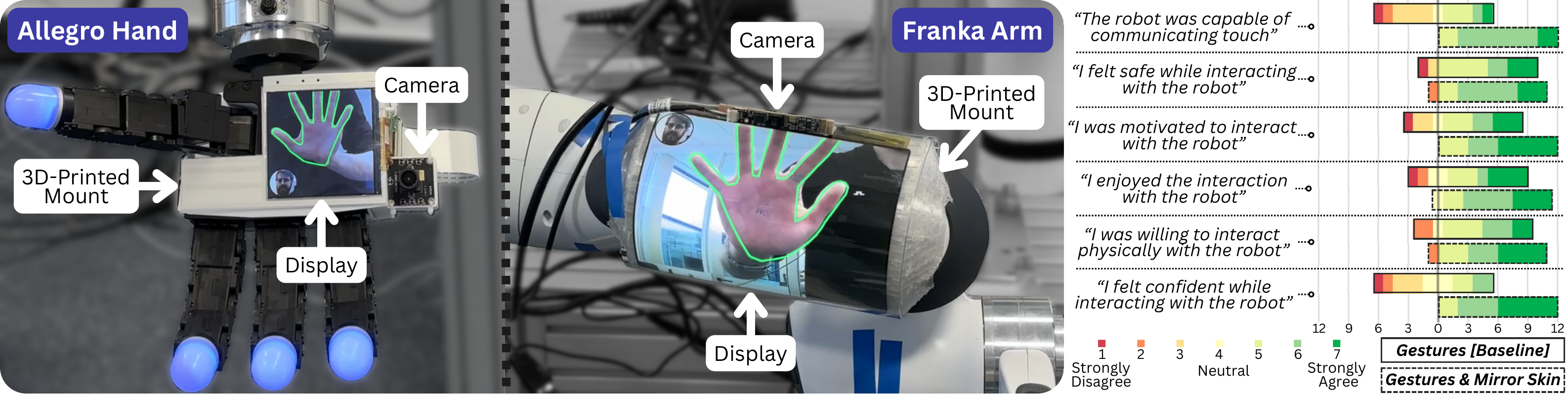}
    \vspace{-0.5cm}
    \caption{We implemented a proof-of-concept prototype of \toolname~on a real robot's forearm and hand (Left) and a user study indicates that our system enhances the user experience compared to only robot gestures (Right).}
    \Description{Left: Two images of our Mirror Skin proof-of-concept prototype that we implemented on an Allegro Hand and a Franka arm. For both the hand and the arm, labels indicate the Mirror Skin main components, that is, a camera, a display and a 3D-printed mount. In both images, the display shows the mirror skin image, targeting a human hand. Right: Stacked bar chart results for 7-point Likert items regarding the robot’s capability to communicate touch intent, the perceived safety, the motivation to interact physically with the robot, the enjoyment during the physical interaction, the willingness to interact physically with the robot and the confidence while interacting physically with the robot. For all items, Mirror Skin + gestures scored better results compared to the gesture baseline. Especially for the capability and confidence, this improvement is large. Further details and analysis are stated in the main text.}
    \label{fig:prototype}
\end{figure*}

\paragraphB{Hardware}
We implemented \toolname~on a humanoid arm-torso setup consisting of two Franka Research FR3 arms from Franka Robotics with seven joints and an Allegro V5 Plus hand from Wonik Robotics with 16 finger joints attached to each arm (cf., \autoref{fig:prototype} (left)). 
For the experiments, only the right arm was used. To capture the human body parts, we used cameras with a wide field of view (FOV) from InnoMaker (1080p; 130°) on the hand and ELP (2448p; 105°) on the arm.
To render the image, we used thin, high-resolution displays from Wisecoco. On the hand, we installed a thin 4-inch LCD (720x720 $\approx$ 255 PPI) and for the arm, we wrapped a flexible 8-inch AMOLED (2460×1860 $\approx$ 385 PPI) around the Franka arm.
We used custom 3D-printed mounts to attach the displays and cameras to the corresponding robot part.

\paragraphB{Software Pipeline}
We implemented the visual mirroring effect in Python, using OpenCV for image capture and processing and Google MediaPipe Landmarks for hand and face tracking. The system includes an initialization procedure that automatically assigns a human actor by detecting their face, capturing a headshot, and storing it as the actor icon.

For each imminent touch interaction, the system allows specification of (1) the robot body part on which the mirror feedback appears, (2) the human body part to be targeted (i.e., left or right hand), and (3) whether the event reflects the robot's intention to touch or an invitation to be touched. Depending on the selected action, the system displays either the robot or human actor icon.

During execution, the system captures the raw camera frame and detects the targeted human body part. Using the MediaPipe Landmarks, the system then creates an outline around the body part by computing a hand-region mask and drawing its contour. Afterward, it determines the image position of the targeted body part $p_{\{\textit{target}\}}$, and computes the translation vector $\vec{v}_{\{\textit{transl}\}}$ required to center the target in the mirror.
To create the dynamic offset, the target is step-wise translated toward the image center at speed $s$:
$$p_{\{\textit{target}\}} = p_{\{\textit{target}\}} + s~\cdot~\frac{\vec{v}_{\{\textit{transl}\}}}{|\vec{v}_{\{\textit{transl}\}}|}$$
Because translating the image introduces empty regions near the camera’s FOV boundaries, we mitigate edge artifacts by cropping a smaller region of the frame and upscaling it to the original resolution through linear interpolation. This creates the illusion of a larger mirror despite the limited FOV. The final frame is then adjusted to the aspect ratio and resolution of the display and rendered as output.

\subsection{Method}

\paragraphB{Experimental design and task}
We implemented three physical interaction scenarios: (1) the robot inviting the human to lean onto the robot's arm for physical support, (2) the robot inviting the participant to place their hand onto the robot’s hand for pulse measurement, and (3) the robot applying a refreshing wipe to the participant’s hand.
Prior to each scenario, participants were informed only about the high‑level interaction goal (e.g., \textit{``the robot aims to measure your pulse''}) but were not told what the physical interaction would look like. Their task was to perform this interaction with the robot based on its \textsc{Feedback}, consisting of \textit{gestures} \textsc{[Baseline]}, or \textit{gestures \& Mirror Skin} [\textsc{Mirror Skin]}. The study used a within-subjects design, i.e., participants experienced all scenarios with each feedback type.
In addition, the robot targeted either the participant’s left or right hand, and its gestures were adapted accordingly to communicate the intended side. Scenario order was counterbalanced using a balanced Latin square, while the \textsc{Feedback} and the target hand were randomized with equal frequency.
To ensure safety, the robot operated in a restricted area and both the experimenter and participant had access to an emergency stop.

\paragraphB{Procedure}
Participants first received a safety briefing and filled out consent and demographic forms. We then explained the \toolname~concept. Subsequently, participants completed all scenarios, and we recorded whether each interaction was performed successfully, i.e., the participant performed the intended physical interaction with the robot.
Afterwards, for each \textsc{Feedback} condition, participants filled out the UEQ-S questionnaire and a custom questionnaire consisting of 7‑point Likert‑scale items (cf., \autoref{fig:prototype} (right)) adapted from prior HRI studies~\cite{UX_in_HRI_Seifi}.
We concluded with a semi-structured interview about participants' preferences and future applications for \toolname.
Each session took approximately 30 minutes and the interview was audio recorded.

\paragraphB{Participants}
We recruited 12 participants (aged 23 to 58, $\bar{x} = 37.67$; 9 male, 3 female; 3 left-handed, 9 right-handed). 5 participants had previously experienced some form of physical interaction with a robot, whereas 7 participants had not.

\paragraphB{Data analysis}
We followed the same steps for data analysis as in study 2 (cf., \autoref{sec:efficiency_study}). For parametric data, we performed paired t-tests and reported Cohen’s $d$ as effect size~\cite{Cohen1988}. 

\subsection{Results}

\paragraphB{Preferences \& User Experience}
Although participants successfully completed the tasks in both conditions (\textsc{Baseline}: $94.4\%$; \textsc{Mirror Skin}: $100.0\%$), most of them (P1-P2, P4-P11) preferred \toolname~over the baseline due to its \textit{``clear and efficient''} (P11) intent communication.
Wilcoxon signed-rank tests showed that they rated the robot's capability to communicate touch intent significantly higher ($p < .05,~r=0.76$) for \textsc{Mirror Skin}~(\med{6}) compared to the \textsc{Baseline} (\med{4}), as shown in \autoref{fig:prototype} (right). Furthermore, participants reported higher willingness ($p < .05,~r=0.70$) to interact physically with the robot for \textsc{Mirror Skin}~(\med{6}) than for the \textsc{Baseline} (\med{5}) and they felt more confident during the physical interaction ($p < .01,~r=0.86$) when \textsc{Mirror Skin}~feedback was provided (\med{7}) compared to the \textsc{Baseline} (\med{4}).
Although participants gave \textsc{Mirror Skin}~high ratings regarding perceived safety (\med{6}), motivation to interact (\med{7}), and enjoyment (\med{6}), these scores did not differ significantly from the \textsc{Baseline} ($p > .05$).

UEQ-S results show that the pragmatic quality of \textsc{Mirror Skin} ($\bar{x} = 1.85$) was significantly higher compared to \textsc{Baseline} ($\bar{x} = -0.33$),$~t(11) =-4.45,~p < .001,~d = -1.285$. Also, hedonic quality scores were significantly higher for \textsc{Mirror Skin} ($\bar{x} = 1.90$) than for \textsc{Baseline} ($\bar{x} = 0.19$), $t(11) =-4.17,~p < .01,~d = -1.205$. As a result, the overall rating of \textsc{Mirror Skin} ($\bar{x} = 1.88$) was significantly higher than \textsc{Baseline} ($\bar{x} = -0.07$), $t(11) =-4.70,~p < .001,~d = -1.359$. Relative to a benchmark dataset of 468 studies~\cite{UEQ-S_Benchmark}, these ratings place \toolname~in the "Excellent" UX category.

\paragraphB{Opportunities \& Applicability}
Participants saw potential for \toolname~across diverse pHRI domains such as medical and caregiving contexts (P1-P3, P6-P11), collaborative and industrial settings (P1, P3-P5, P8-P9, P11-P12), educational scenarios (P2-P3, P10, P12), and home or service robots (P6, P8-P9, P11).

Especially in collaborative scenarios like object handovers, participants (P1, P4, P5, P8, P11) saw opportunities for rich semantic feedback, as current gestural interactions are not precise enough. P4 noted: \textit{``If [the robot] tries to hand over something [...] you put two hands up to see which hand [the robot] wants to touch''}. \toolname\textit{'s}~rich semantic feedback, which tells exactly where contact happens, would reduce ambiguity in these scenarios.

Furthermore, participants indicated that \toolname~is particularly suited for initial interactions with robots, as the feedback was found to be \textit{``very intuitive''} (P1) and \textit{``exciting''} (P5), which encourages humans to interact physically with the robot. P8 noted that this may be particularly beneficial for robot-assisted medical treatments, for example \textit{``in physiotherapy, during the massage, it gives you more security or more confidence, because you have confirmation that the robot aims for the right body part''}.

Moreover, P3 hypothesized that \toolname~is particularly important for robots with non-humanoid morphology, as \textit{``it's easier to communicate with \toolname, 'touch this!' than [verbally describe] whatever is between the third and the fourth joint''}, thus indicating advantages of rich visual feedback over other modalities.
Finally, participants (P5, P8, P11, P12) suggested extending the mirror metaphor to signal points of interest, for example, communicating the objects a robot plans to grasp.
\section{Discussion \& Future Directions}
In this section, we interpret our findings and outline implications for advancing visual skin‑based communication. We also discuss the limitations of this work and identify directions for future research.

\subsection{Enhancing Robot Touch Intent Communication with \toolname}
Our results demonstrate that \toolname~enables faster and more accurate interpretation of touch intent compared to gestures and gaze.
One key advantage of visual, skin‑based communication is its ability to convey all relevant information at a glance, without requiring robot motion. This enables users to anticipate forthcoming actions and thus could be an efficient mechanism for safe pHRI.

We also found that gesture‑based communication (e.g., pointing) offers limited spatial precision, particularly when the robot’s end effector is farther away from the human. In contrast, \toolname~provides fine‑grained spatial feedback, which is well‑suited for conveying precise touch points.
Furthermore, unlike other feedback methods (e.g., gaze) which are typically limited to a single point of focus, \toolname~supports the simultaneous visualization of multiple touch intentions. Future work should investigate this spatial multiplexing as it offers interesting opportunities for communicating touch intent in multi-contact or multi-user scenarios.

Finally, we showed that clearer articulation of the robot’s intent increases humans’ willingness and confidence to engage in physical contact, highlighting the value of higher-semantic feedback. As robots increasingly perform physical interactions in domains such as healthcare and industry collaboration, fostering human acceptance of robot touch becomes crucial.

\subsection{Expanding \toolname\textit{'s} Semantic Information for Robot Touch}
Similar to the visual feedback of cephalopods, \toolname~is not constrained in what information can be visually mapped onto the robot’s skin.
A natural extension of communicating touch actions is to explore how \toolname~can be leveraged to convey information about the quality of the touch, such as the robot’s precise touch action and the intensity of contact. We hypothesize that providing humans with anticipatory information about how the robot will touch their body may enhance comfort and trust.

Another promising direction is to extend \toolname~to support bidirectional communication, enabling the robot to respond to human touch intent. For instance, when a human hand approaches the robot’s arm, \toolname~could be used to signal the robot’s awareness of the incoming interaction and visually indicate acceptance, rejection, or suggest a counterproposal. 
When a human hand approaches a sensitive or safety-critical region of the robot’s body, \toolname~could be employed to deny the interaction and redirect it to a safer adjacent location. Given the high spatial accuracy and rapid interpretability of \toolname, such responsive feedback could facilitate real-time negotiation of touch interactions, enhancing both safety and fluidity in human-robot collaboration.

\subsection{\toolname~for Diverse Robot Communication and Morphologies}
While our work focused specifically on touch intent, the underlying principles do apply to other forms of human-robot communication. For instance, building on the work of Krüger et al. \cite{MirrorEyes_Human, MirrorEyes_Object}, \toolname~could be extended to support spatial referencing, such as indicating the robot’s intention to grasp a specific object. Crucially, unlike prior mirroring approaches that centralize feedback to one specific region (e.g., the robot’s eyes), \toolname~localizes the visual feedback directly at the relevant body part. This eliminates the need for humans to divide attention between multiple regions (e.g., gaze and end effector), potentially increasing efficiency (cf., \cite{Light_RLS}).

Moreover, we see potential in high-resolution, skin-based visual feedback, including textures, images and videos, as a promising expressive medium for robots. Building on the work of Hu et al. \cite{ShapeChange_Exploration}, such visual feedback could convey messages to the human, indicate internal robot state, or communicate emotional affect during social human-robot interactions.

Furthermore, to simplify the interpretation of gestures and gaze, we restricted touch actions to be performed with the hand. However, in real-world scenarios, robots may initiate contact using other body parts, for example, lifting a person with the forearm or stabilizing with the torso. \toolname~appears to be well-suited to communicate such interactions, as it is inherently morphology-independent. This also offers a promising communication modality for appearance-constrained robots (e.g., \cite{AR_Drone, Cooperation_Drones, Light_Bioluminescence, Shape_GooY}) that lack anthropomorphic features and hence cannot rely on conventional nonverbal cues such as gestures or gaze.

Finally, although we attached all screens for \toolname, numerous existing robotic platforms already incorporate body-mounted displays, particularly on the torso \cite{Screens_Body} and face \cite{Screens_Face}. These existing displays could be directly repurposed for \toolname, or supplemented with additional screens to extend coverage. As skin-based display technology matures, we envision the entire robot body eventually functioning as a continuous display surface (cf., \cite{DisplaySkin}).

\subsection{Limitations \& Future Work}

While \toolname~demonstrates strong potential for communicating touch intent, our current implementation is subject to several limitations that we plan to address in future work. 

\paragraphB{Advancing the \toolname~prototype}

In our current implementation of \toolname, self-occlusion remains an open challenge, particularly in dynamic scenarios where the robot's own body may obstruct the visual feedback. In future work, we intend to address this limitation through several approaches. First, occlusion effects could be mitigated by rendering \toolname~on adjacent, non-occluded surface regions while maintaining a visual reference to the originally targeted surface area. Moreover, backwards-facing cameras could be integrated to simulate visual transparency of occluding body parts (cf., \cite{TransparentCar}). Finally, supplementary hardware such as AR headsets could be employed to render visual feedback directly above or through occluding body parts (cf., \cite{AR_Arrow}).

At the same time, our proof-of-concept was limited to only one hand and forearm of the robot. In addition, although we used wide-angle cameras, they did not cover every angle during robot motion. Future work should equip a robot with more displays and cameras to remove blind spots and evaluate \toolname~on a fully scaled robot.

\paragraphB{Tailor \toolname~to dedicated settings and scenarios}
First, the current design of \toolname~has been evaluated with able-bodied adults. However, key application domains such as caregiving often involve user groups with varying perceptual and cognitive profiles, including individuals with visual impairments. Consequently, future work should conduct a stakeholder analysis to identify the specific needs of these user groups and explore adaptations of the concept to ensure inclusive communication.

Second, to remain within a feasible scope, we evaluated \toolname~only in a controlled, safe lab environment. 
Future work should investigate its performance and cognitive load across a broader range of contexts and tasks, and adapt \toolname\textit{'s} visual feedback accordingly. For example, during repetitive or recurrent interactions, the feedback could be simplified and made more static (e.g., by removing depth and dynamic offset) to reduce visual load. Additionally, although we compared \toolname~with gestures and gaze as established non-verbal communication modalities, future research should explore its integration into multimodal robotic systems to potentially further advance communicative efficiency.

Third, our findings suggest that \toolname~is particularly well-suited to initial human-robot interactions, as it appears to encourage physical engagement with the robot. A next step will be to investigate the effects of \toolname~during long-term use, once the potential novelty effect has diminished. To sustain engagement and efficiency over long-term use, it may be advantageous to grant users control over customizing the visual feedback along our proposed design dimensions.

Fourth, \toolname~requires live camera images of people and the environment, which could raise privacy concerns for robots in public spaces, especially regarding bystanders who are not actively interacting with the robot. Therefore, we emphasize techniques like background removal and face blurring to mitigate these issues and encourage further investigation in future work.  

Finally, the potential intuition of the \toolname~deserves further investigation. The mirror metaphor appears to facilitate the understanding and quick use of the displayed information. We observed that participants actively utilized laws governing the use of actual mirrors for disambiguation, e.g., by moving their wrist or fingers to identify the targeted hand. However, our evaluation of \toolname~was conducted in scenarios where humans were explicitly introduced to the interface concept and its semantics. Future work should investigate \toolname\textit{'s}~use without such guidance and, if required, identify further means for promoting intuitive use.

\section{Conclusion}

In this work, we introduce \toolname, a cephalopod-inspired concept for conveying robot touch intent via high-resolution, mirror-like visual feedback on robot skin. By mapping visual representations of human body parts onto the robot’s touching surface, \toolname~communicates \textit{who} shall initiate touch, \textit{where} it will occur and \textit{when} it is imminent.
We developed and evaluated the \toolname~concept through a mixed methods approach consisting of a structured design exploration in VR, a proof-of-concept prototype and three controlled user studies. Our results demonstrate that \toolname~significantly improves both accuracy and response time in interpreting touch intent compared to a gesture-and-gaze baseline and that \toolname~enhances the UX, including greater willingness and confidence to engage in pHRI.
\toolname~offers new avenues for robot skin-based visual interfaces, dynamic bidirectional human-robot interaction, multi-target touch communication, and morphology-independent feedback. This makes \toolname~a promising concept for various robotic platforms and applications.

\begin{acks}
We thank all participants of our user studies, as well as Oliver Schön, for his support during the fabrication of the \toolname~prototype. Moreover, we thank Hannah Wagmann for sketching our conceptual space. We also thank the reviewers for their valuable comments.
\end{acks}

\bibliographystyle{ACM-Reference-Format}
\bibliography{bibliography}

\appendix
\newpage
\section{Performance Study} \label{sec:appendix_feedback}

To give a better impression of the VR intent communication (cf., \autoref{sec:efficiency_study}), we provide example images of the robot conveying an invitation to be touched by the human, and its intention to touch the human, for \textsc{Baseline} (gesture \& gaze), \textsc{Pre-motion Mirror Skin}, and \textsc{In-motion Mirror Skin} (\autoref{fig:appendix_feedback}):

\begin{figure} [H]
    \centering
    \vspace{-0.3cm}
    \includegraphics[width=0.75\linewidth]{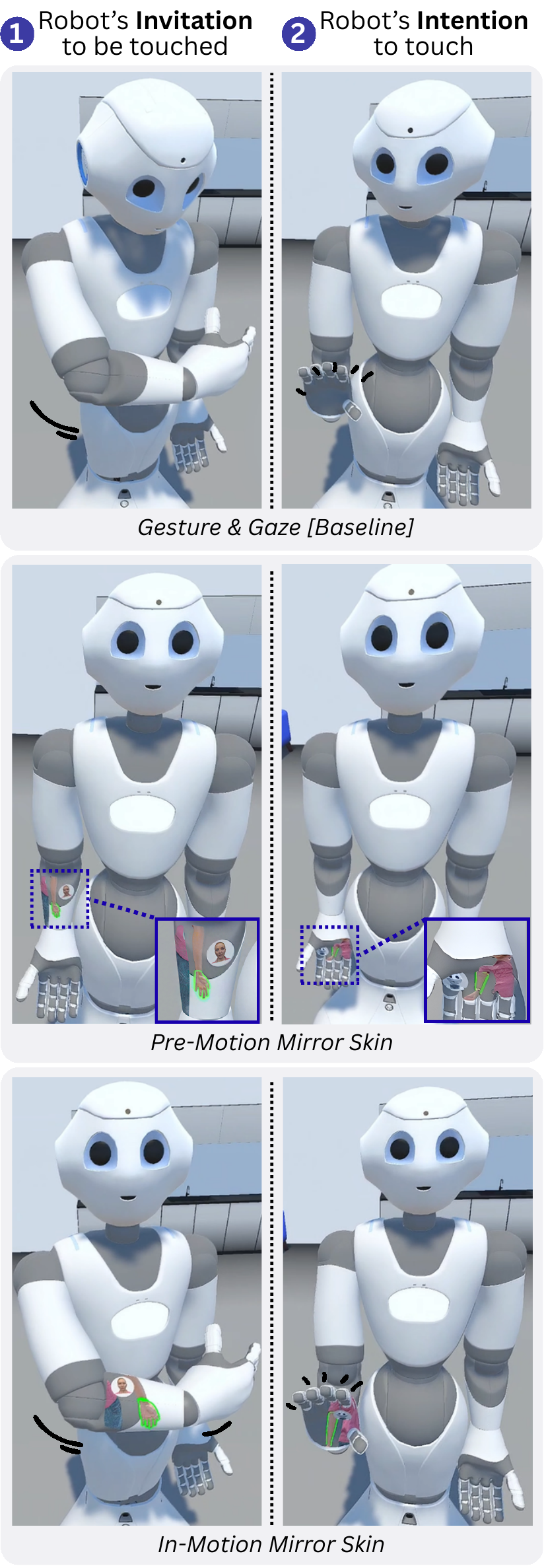}
    \vspace{-0.4cm}
    \caption{Examples of the robot communicating (1) an invitation to be touched, or (2) its intention to touch the human, for all three feedback techniques of study 2.}
    \Description{Shows six images from the controlled VR study setup. For each feedback type, that is, gestures and gaze, pre-motion Mirror Skin and in-motion Mirror Skin, the figure depicts one example of a humanoid robot inviting the human to touch it or the robot’s intention to touch the human. All images are from the participant’s first-person view in VR. 1) Gesture and gaze invitation: The robot gazes in the direction of the participant's right hand, while offering the robot’s right forearm to the participant to communicate that the robot wants to be touched there by the participant’s right hand. 2) Gesture and gaze intention: The robot gazes in the direction of the participant’s left forearm while stretching out its right hand to touch this body part, thereby communicating its intention to touch the left forearm of the participant with its right hand. 3) Pre-motion Mirror Skin invitation: The robot stands in a neutral pose (both arms casually at the side) and mirrors the human's right hand onto its own right forearm. That way, the robot communicates that it wants to be touched there by the participant’s right hand. 4) Pre-motion Mirror Skin intention: The robot stands in a neutral pose (both arms casually at the side) and mirrors the human’s left forearm onto its own right hand. Thus, the robot conveys that it wants to touch the human’s left forearm with its right hand. 5) In-motion Mirror Skin invitation: The robot has the same arm pose as in the gesture and gaze invitation and the same Mirror Skin feedback as in the pre-motion Mirror Skin invitation image and communicates the same message. 6) In-motion Mirror Skin intention: The robot has the same arm pose as in the gesture and gaze intention and the same Mirror Skin feedback as in the pre-motion Mirror Skin intention image and communicates the same message.
}
    \label{fig:appendix_feedback}
\end{figure}

\end{document}